# The Proximity Effect of QSO Metal Line Absorption Systems on Lyman $\alpha$ Forests

Xiangdong Shi
*Department of Physics, Queen's University, Kingston, Ontario, Canada, K7L 3N6*

**ABSTRACT**
The proximity effect of metal line absorption systems in spectra of high redshift quasars and quasar pairs is investigated through the distribution of Ly$\alpha$ forests. No strong proximity effect is found down to 500 km/sec away from a metal line absorption system for lyman $\alpha$ clouds with $13.3 \leq \log N_{\rm HI} < 13.8$ at a redshift of $\sim 3$. Its implication on properties of high redshift galaxies and Ly$\alpha$ clouds is discussed. More high resolution spectra of high redshift quasars and pairs of quasars (whose lines of sight are separated by $\sim 0.3 h^{-1}$ to $1 h^{-1}$ Mpc) are needed so that the proximity effect can be reliably determined for Ly$\alpha$ clouds with various $N_{\rm HI}$ and down to a closer distance.

## 1 INTRODUCTION

The metal absorption lines and lyman $\alpha$ forests in quasar spectra are unique probes of structure of the universe at high redshifts. Their origin and properties have been investigated intensively both observationally and theoretically. There is compelling evidence that metal line absorptions arise from galactic halos of luminous galaxies (see, for example, Sargent, Boksenberg and Steidel 1988, hereafter 1988; Bergeron 1988; Crotts et al. 1994; and references therein), while lyman $\alpha$ lines are believed to originate from highly ionized hydrogen clouds in the intergalactic medium (see review of Bajtlik 1993). The detailed modeling of these galaxies and clouds at high redshift, however, proves to be challenging and sometimes controversial. How these metal line absorption systems are are distributed in space, and how their star formation proceed, are still subject to further investigation. For Ly$\alpha$ clouds, the major controversy is its confining mechanism: whether they are confined by pressure of a hot intergalactic medium (Sargeant et al. 1980; Ikeuchi and Ostriker 1986) or gravity from the mini-halo of dark matter in which they are embedded (Rees 1986; Ikeuchi 1986), or by ram-pressure of shocks of infalling gas (Cen et al. 1994), or a hybrid of them (Bond, Szalay & Silk 1986; Salpeter 1993; Meiksin 1994; Cen et al. 1994).

The physical properties of these high redshift objects are very important to our understanding of the universe as a whole. For example, their sizes, masses, correlation, ionization state, etc., can reveal a great deal of information about the underlying cosmology model. In this letter, we attempt to explore how these two populations of high redshift objects interact with each other, by probing the proximity effect of metal line absorption systems in quasar spectra through the distribution of Ly$\alpha$ forests. The proximity effect of quasars in their own spectra has proven to be a very useful tool to obtain knowledge about the background ionization photon flux (Bajtlik, Duncan & Ostriker, 1988; Lu, Wolfe & Turnshek 1991). We hope the proximity effect of metal line absorption systems (i.e. galaxies) can also provide useful information about our universe at high redshift. A similar problem–the cross correlation between metal line absorption systems and Ly$\alpha$ clouds–has been investigated by Barcons and Webb (1990). The correlation between Ly$\alpha$ clouds and galaxies at $z \lesssim 0.2$ is also explored toward the line of sight of 3C273 (Morris et al. 1993). We will explore this correlation at close distance with higher resolution data at a redshift of $\sim 2$ to 3, and try to use cleaner ways to remove the contamination from Ly$\alpha$ absorbtion of galaxies in our Ly$\alpha$ sample, as discussed below.

Several groups have recently done numerical simulations of formation of Ly$\alpha$ clouds (Cen et al. 1994; Zhang et al. 1995; Mücket et al. 1995; Hernquist et al. 1995). A general feature arising from these simulations is that Ly$\alpha$ clouds with higher HI column density ($N_{\rm HI} \gtrsim 10^{13}$ cm$^{-2}$) usually trace dark matter and form in overdense region of the universe, with clouds having $N_{\rm HI} \gtrsim 10^{14}$–$10^{15}$ cm$^{-2}$ being often isolated and clouds having $\log N_{\rm HI} \lesssim 10^{14}$cm$^{-2}$ being more diffusive in form of sheets or filaments. Galaxies, which provide metal line absorption in quasar spectra, also form in the overdense region. Therefore, galaxies and Ly$\alpha$ clouds can form with proximity in overdense environment. This feature is expected in all model where Ly$\alpha$ clouds are at least partially confined by potential well of dark mat-ter. Thus an investigation of proximity effect is possible and sensible.

**Table 1.** Our QSO spectra sample.

| QSO | data range | Metal system |
| --- | --- | --- |
| 0055-269[a] | 4824Å–5567Å | $z_{\rm abs} = 3.086$ |
|  |  | $z_{\rm abs} = 3.19$ |
|  |  | $z_{\rm abs} = 3.256$ |
|  |  | $z_{\rm abs} = 3.437$ |
| 0636+680[b] | 4315Å–4894Å | $z_{\rm abs} = 2.904$ |
| 0014+813[c] | 4500Å–5237Å | $z_{\rm abs} = 2.80$ |
|  |  | $z_{\rm abs} = 3.227$ |
| 1946+7658[d] | 4256Å–4843Å | $z_{\rm abs} = 2.644$ |
|  |  | $z_{\rm abs} = 2.844$ |
|  |  | $z_{\rm abs} = 2.893$ |

[a] Cristiani et al. 1995;
[b] Hu et al. 1995;
[c] Rauch et al. 1992;
[d] X.-M. Fan and D. Tytler, 1994.

Besides the gravitational interaction among Ly$\alpha$ clouds, galaxies, and their underlying dark matter potential wells, the ionization photon flux from star burst in galaxies may also play an important role in the proximity effect, if galaxies contribute significantly to the background ionization photon flux. Other factors, such as pressure from inter-galactic medium or accretion shocks, should also be considered.

In section 2, we will investigate the distribution of Ly$\alpha$ absorption lines in the neighborhood of metal line absorption systems. Since we know that metal line absorption systems are galaxies that also produce Ly$\alpha$ absorptions usually in the form of Damped Ly$\alpha$ lines and Lyman limit systems, with a high column density $N_{\rm HI} \gtrsim 10^{15.5} {\rm cm}^{-2}$, we divide our Ly$\alpha$ forest sample into two groups: those with $\log N_{\rm HI} \geq 13.8$, and those with $13.3 \leq \log N_{\rm HI} < 13.8$. At least those with the lower column density should be free of contamination of Ly$\alpha$ absorptions in galaxies. This is also consistent with the lack of metal detection in systems with $N_{\rm HI} < 10^{14}$ cm$^{-2}$ (Tytler & Fan 1994; Cowie et al. 1995). We will also investigate the same distribution of Ly$\alpha$ absorption lines in spectra of quasar pairs whose lines of sight are separated by a physical distance of $\sim 0.3h^{-1}$ to $1h^{-1}$ Mpc. Since the metal line absorption systems are not large enough to span both lines of sight, this clearly remove any mixing between Ly$\alpha$ absorption from Ly$\alpha$ clouds and that from galaxies. In section 3, we will discuss the implication of our result in section 2, and suggest future observations that may extend the investigation of the proximity effect of metal line absorption systems to a higher statistical level.

## 2  PROXIMITY EFFECT OF METAL LINE ABSORPTION SYSTEMS

Table 1 lists the data used in our first test. We choose these data because they are from recent high resolution observations in roughly the same redshift range of $z \sim 3$, and are complete in Ly$\alpha$ absorption lines down to $N_{\rm HI} \geq 10^{13.3}$ cm$^{-2}$ and $b \geq 14$ km/sec (where $b$ is the doppler parameter of an absorption line), which is also our selection criterion of Ly$\alpha$ samples. To avoid the proximity effect of quasars, we excludes lines within 5000 km/sec of the quasar redshifts. We also excluded lines shortward of the Lyman $\beta$ emission of the quasars. We group Ly$\alpha$ absorption lines in the vicinity of a metal line absorption system into 3 bins: $\Delta V < 500$ km/sec where $\Delta V$ is the velocity split between a Ly$\alpha$ absorber and the closest metal line absorption system (bin I); 500km/sec $< \Delta V <$ 1000 km/sec (bin II); 1000km/sec $< \Delta V <$ 1500 km/sec (bin III). Since the Hubble expansion rate at $z \sim 3$ in the standard $\Lambda = 0, \Omega = 1$ cosmology is $(1+z)^{1.5} H_0 \sim 800h$ km/sec/Mpc, the second bin roughly corresponds to a physical distance of $\sim 1h^{-1}$ Mpc, when peculiar velocities are neglected. (In a $\Lambda = 0.7, \Omega = 0.3$ universe the Hubble expansion is $\approx 440h$ km/sec/Mpc at $z = 3$. But to yield a similar age, $h$ in this model has to be 45% higher than that adopted in the $\Lambda = 0, \Omega = 1$ universe. Thus the distance scale in question differs only by 20% in the two models when a similar age is assumed. We will only quote the distance scale in the standard $\Lambda = 0, \Omega = 1$ universe.) Our result is summarized in Table 2.

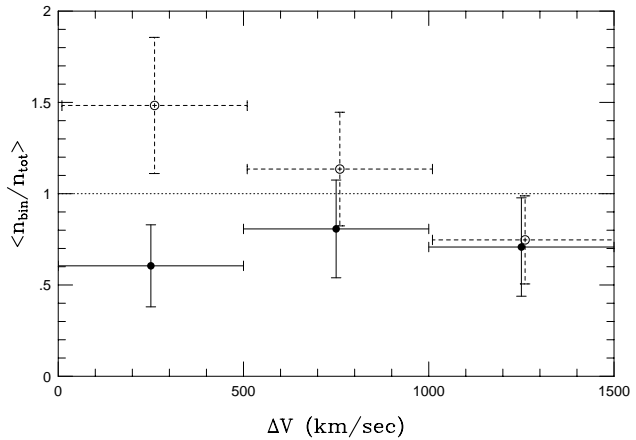

**Figure 1.** The number density of Ly$\alpha$ absorption lines in the neighborhood of a metal absorption line system in unit of the average Ly$\alpha$ line density, vs. the velocity split between Ly$\alpha$ lines and the metal system. Filled circles: $10^{13.3} {\rm cm}^{-2} \leq N_{\rm HI} < 10^{13.8} {\rm cm}^{-2}$; open circles: $N_{\rm HI} \geq 10^{13.8} {\rm cm}^{-2}$.

Assuming that the errors in the numbers of lines are their square roots, Figure 1 shows the weighted average of $n_{\rm bin}/n_{\rm tot}$ (where $n_{\rm bin}$ is the average line density in each bin in a quasar spectrum and $n_{\rm tot}$ is the overall line density of the spectrum) as a function of $\Delta V$. For simplicity, no line density evolution is assumed. This will not affect our following discussion significantly due to the small redshift range investigated in each quasar. Even though the errorbars in Figure 1 are large because of the small number of quasars we have investigated, the figure nonetheless suggests that the distribution of Ly$\alpha$ clouds with a low HI column density ($13.3 \leq \log N_{\rm HI} < 13.8$) near a metal line absorption system is consistent with a uniform distribution or a weak (anti-)correlation around the metal system; while data of Ly$\alpha$ clouds with a higher HI column density ($\log N_{\rm HI} \geq 13.8$) may suggest a positive correlation, which is not surprising since the sample contains metal line absorption systems that are known to be able to produce multiple absorptions and cluster strongly at close distances (SBS).

**Table 2.** Number of Lyα lines with $13.3 \leq \log N_{\rm HI} < 13.8$ [$\log N_{\rm HI} \geq 13.8$] that satisfies our criterion in our sample. I: $\Delta V < 500$ km/sec; II: $500$ km/sec $< \Delta V < 1000$ km/sec; III: $1000$ km/sec $< \Delta V < 1500$ km/sec.

| QSO | Total # of lines | # of lines in I | # of lines in II | # of lines in III |
|---|---|---|---|---|
| 0055-269 | 86 [115] | 4 [21] | 8 [12] | 7 [8] |
| 0636+680 | 69 [68] | 2 [2] | 4 [2] | 0 [1] |
| 0014+813 | 87 [92] | 2 [4] | 2 [4] | 4 [4] |
| 1946+7658 | 49 [36] | 4 [8] | 3 [5] | 4 [2] |

To quantify the above assertions, we perform a fit using Poisson statistics, assuming the following cross correlation for Lyα clouds around a metal line absorption system

$$\xi_{\rm ML}(\Delta V) = \max[-1, \beta \left(\frac{\Delta V}{500\,{\rm km/sec}}\right)^{-\alpha}], \quad \text{with } \alpha \leq 3, \quad (1)$$

where $\alpha$ and $\beta$ are fitting parameters. $\beta$ describes the amplitude of the cross correlation as well as the sign of it (whether it is a correlation or anti-correlation). $\alpha$ denotes how fast the cross correlation falls off with the distance. It is restricted to be no larger than 3 due to the following two considerations: first, anti-correlations with very large $\alpha$ essentially behave similarly, that is, to make the expected number of lines in bin I approaching 0; while positive correlation with very large $\alpha$ is clearly disfavored by data. Secondly, linear theory of gravitational clustering yields a correlation function $\propto r^{-3-n}$ if the underlying power spectrum $\propto k^n$ (where $k$ is the wave number). As we know $n$ is negative at the scale of our concern.

Figure 2 shows the allowed parameter space at 95% C.L. for the fit. Several conclusions may be drawn from the figure: (1) Both data set, especially the data with the lower $N_{\rm HI}$, are consistent with zero cross correlation with metal line absorption systems. (2) For the lower $N_{\rm HI}$ data in Fig. 2(a), $|\beta| \ll 1$ when $\alpha \lesssim 1$, and $\beta$ lies between $-1$ and $0$ when $\alpha$ is large (where the statistics is dominated by bin I). This suggests that there is no strong correlation or anti-correlation with metal line absorption systems extending beyond the first velocity bin ($\Delta V \gtrsim 500$ km/sec) in the $13.3 \leq \log N_{\rm HI} < 13.8$ Lyα clouds. While a strong positive correlation is highly disfavored, an anti-correlation at small distance (with $\Delta V \lesssim 500$ km/sec) is possible and in fact favored because of the smaller line density in bin I. (3) In Fig. 2(b), the data show evidence of the higher $N_{\rm HI}$ Lyα clouds clustering around metal line absorption systems, which as we discussed before is not surprising because they contain Lyα lines arising from galaxies. In fact, even if we assume conservatively that each metal line absorption system contributes only one Lyα absorption with $N_{\rm HI} \geq 10^{13.8}$ cm$^{-2}$ and we exclude them from our sample, the number of Lyα lines with $N_{\rm HI} \geq 10^{13.8}$ cm$^{-2}$ in bin I will be reduced by $\sim 1/5$ to $1/2$ in each quasar spectrum, essentially wiping out any evidence for clustering around galaxies. Our finding that there is no strong evidence for significant proximity effect near metal line absorption systems down to $\Delta V \sim 500$ km/sec is in general agreement with that of Barcons and Webb (1990), although they reached their conclusion based on a lower resolution sample including all Lyα lines with a rest equivalent width larger than $0.36$Å($N_{\rm HI} \gtrsim 10^{14}$cm$^{-2}$). By dividing our sample into two ranges of $N_{\rm HI}$, we can clearly remove any possible contamination from metal line absorption systems in the lower $N_{\rm HI}$ Lyα sample.

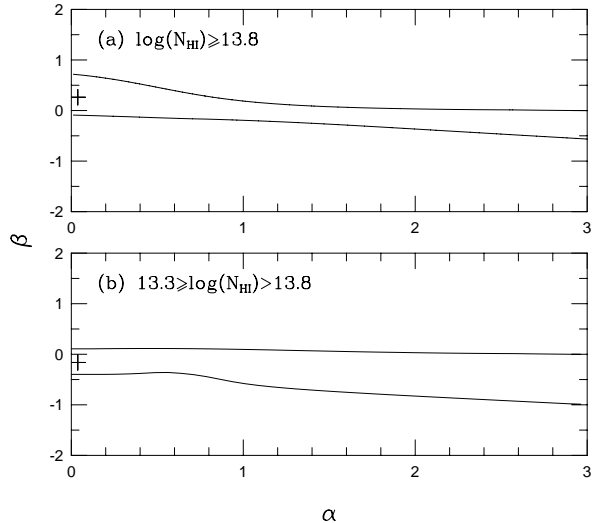

**Figure 2.** Assuming a cross correlation for Lyα absorption lines around a metal line absorption system of the form $\xi_{\rm ML} = \max[-1, \beta \Delta V_{500}^{\alpha}]$, regions within contours are allowed at 95% C.L. Crosses are the best fits.

We must caution, however, that the current data set is still not large enough to make compelling conclusions, and that more high resolution data at $z \sim 3$ are essential to provide better statistics and to explore the proximity effect reliably down to a lower velocity split.

A second and complementary way to examine the proximity effect of metal line absorption systems is to examine spectra of quasar pairs whose lines of sight are separated by $\sim 0.3h^{-1}$ to $1h^{-1}$ Mpc. This separation is close enough to be interesting, but large enough to remove the contamination in the Lyα sample by Lyα lines from galaxies, because the size of a typical metal line absorber is $\sim 100h^{-1}$ kpc (SBS) and the size of today's galactic halo can extend to $\sim 200$ kpc (see review of Ashman 1992). Quasar group 1628+2651A/B and 1638+2653 (Crotts 1989) provides a convenient example for this method. The three quasars in the group are separated by $147''$ (1628+2651A and 1628+2653), $127''$ (1628+2651A and B) and $177''$ (1628+2653 and 1628+2651B) respectively. This separation roughly corresponds to $0.5h^{-1}$–$0.7h^{-1}$ Mpc at a redshift of 2 for $0.3 \leq \Omega \leq 1$ cosmology. By computing the number density of Lyα absorption lines within the neighborhood of a metal line absorption system in a neighbor quasar spectrum, and comparing it with the average Lyα line density in the first line of sight, we can determine the magnitude of the proximity effect. Table 3 summarizes our computation from Crotts' (1989) data of the quasar trio. The spectrum range under investigation is similarly selected as in Table 1. Only Lyα lines with an observed equivalent width larger than $0.2$Å($N_{\rm HI} \gtrsim 10^{13.3}$ cm$^{-2}$), which are believed to be complete, are counted. We choose to count Lyα lines within $\lesssim 1h^{-1}$ Mpc of a metal line absorption system. Again evolution of the line density is ignored.

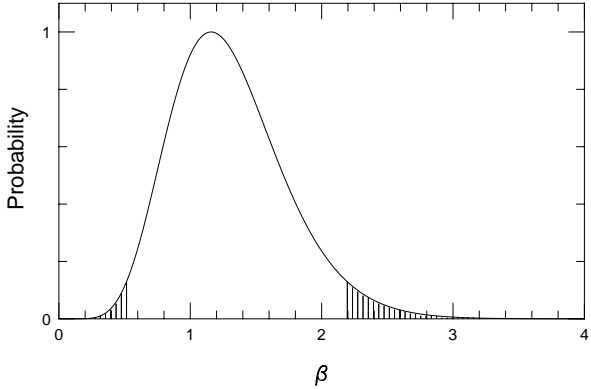

**Figure 3.** Probability distribution of parameter $\beta$, the ratio of the local number density of Ly$\alpha$ absorption lines in the neighborhood of a metal line absorption system, to the average Ly$\alpha$ line density, in spectra of quasar trio 1628+2651A/B and 1628+2653. Shaded regions are excluded at 95% C.L.

Since there is only one velocity bin, we fit the data assuming

$$\langle n_L \rangle = \beta \langle n_{tot} \rangle \qquad (2)$$

where $\langle n_L \rangle$ is the local Ly$\alpha$ line density within a line of sight range of $\pm 0.8h^{-1}$ Mpc centered at the redshift of a metal line absorption system in a neighbor line of sight. $\langle n_{tot} \rangle$ is the average Ly$\alpha$ line density in the entire spectrum range considered in the first line of sight, and $\beta$ the free parameter describing the strength and sign of the proximity effect. Figure 3 shows the probability distribution of $\beta$ from the data set using Poisson statistics. At 95% C.L., $\beta$ is constrained to be within 0.5 to 2.2, consistent with zero or moderate correlation (or anti-correlation) at a physical distance of $\sim 1h^{-1}$ Mpc from metal line absorption systems at $z \sim 2$. Of course, the statistics in available data is very low, and not all the metal line absorption systems in the data set are identified with high confidence (Crotts 1989). But this method of looking for quasar pairs whose lines of sight are separated by $0.3h^{-1}$ Mpc $\lesssim r < 1h^{-1}$ Mpc may prove very powerful in investigating the relation between Ly$\alpha$ clouds and galaxies, because the Ly$\alpha$ sample is free of contamination from galaxies. More data from such quasar pairs, especially at high $z$, are highly desirable.

## 3 DISCUSSION

As shown in the previous section, Ly$\alpha$ lines with $10^{13.3}$ cm$^{-2} \leq N_{HI} < 10^{13.8}$ cm$^{-2}$ at $z \sim 3$ suggest that there is no significant proximity effect of galaxies in these Ly$\alpha$ clouds down to a velocity split of 500 km/sec to 1000 km/sec from metal line absorption systems at $z \sim 3$. Although the statistics needs improving, we nonetheless attempt to explore what the implications are on structure formation at high redshift if such an assertion is true.

Metal line absorption systems, thus galaxies, were shown to cluster strongly at high redshift (SBS). CIV absorption lines at $\langle z \rangle \sim 2$ show that the average amplitude of correlation function within $\Delta V \leq 600$ km/sec is 6–11 (most of it cannot be attributed to multiple absorption from a single galaxy because pairs of lines with splits below 200 km/sec are excluded) (SBS). The average number of galaxies within $1h^{-1}$ Mpc of a detected galaxy is

$$N(< 1h^{-1} \text{Mpc}) = \int_0^{1h^{-1} \text{Mpc}} 4\pi r^2 n_{ML}(z)[1 + \xi(r)]dr \qquad (3)$$

where $n_{ML}$ is the number density of the absorbers and $\xi(r)$ is their auto-correlation function. If these metal line absorption systems are normal galaxies with $n_{ML} \approx 0.014h^3(1+z)^3$ Mpc$^3$ (Loveday et al. 1992), applying this equation to metal line absorption samples of SBS yields that at $z \sim 2$, there will be on average 11 more galaxies within $1h^{-1}$ Mpc of a detected galaxy. Extrapolating this clustering to $z \sim 3$ yields that on average 14 galaxies will lie within $1h^{-1}$ Mpc of one detected galaxy, with a number density 4 times the average number density of galaxies. The underlying dark matter potential well of the group of galaxies should therefore be fairly deep. The deep potential well tends to accrete gas, including Ly$\alpha$ clouds around them, regardless of the detail of forming Ly$\alpha$ clouds. On the other hand, other factors, such as tidal force, ionization photon flux from star bursts in galaxies may destroy these clouds or reduce their HI column density, and pressure in intergalactic medium or from shock waves may prevent them from falling into the dark matter potential well. We will discuss each of these factors at a velocity split of 500 km/sec $< \Delta V < 1000$ km/sec, corresponding to a physical distance of $\sim 1h^{-1}$ Mpc at $z \sim 3$.

**(1) Accretion.** In linear theory, assuming the simplest spherical geometry, the peculiar velocity field $v_{pec}$ at a distance $R$ from origin in comparison to the Hubble expansion $HR$ is (Shi, Widrow & Dursi 1995)

$$\frac{v_{pec}}{HR} \sim -\frac{2}{3}\left(\frac{\delta M}{M}\right)_R \qquad (4)$$

where $(\delta M/M)_R$ is the mass fluctuation within radius $R$. Thus for Ly$\alpha$ clouds around galaxies, if the absolute value of this ratio is comparable to unity, it means that the accretion of Ly$\alpha$ clouds into the gravitational potential well will be significant within a Hubble time of $H^{-1}$ at a distance $R$. Barring other factors that may affect the clouds, strong clustering of Ly$\alpha$ clouds should show within $R$ of the galaxies. The clustering will be even stronger in redshift space, as in quasar spectra. By requiring no strong accretion of Ly$\alpha$ clouds around galaxies within $1h^{-1}$ Mpc at $z \sim 3$, we have

$$\left(\frac{\delta M}{M}\right)_{1h^{-1}\text{Mpc}} \lesssim 1 \quad \text{at } z \sim 3 \qquad (5)$$

where $1h^{-1}$ Mpc represents physical distance, corresponding to a comoving distance of $4h^{-1}$ Mpc. Standard Cold Dark Matter model with $\Omega = 1$, $\Lambda = 0$ and $H_0 = 50$ km/sec/Mpc normalized by COBE predicts a $\langle(\delta M/M)^2\rangle^{1/2} \approx 0.6$ at this scale at a redshift of 3 (Efsthathio, Bond & White 1992), consistent with eq. (5). (In contrast, such a model predict $\langle(\delta M/M)^2\rangle^{1/2} \approx 0.9$ at $2h^{-1}$ Mpc comoving scale at $z = 3$.) Given an average local galaxy number density within $1h^{-1}$ Mpc of a detected galaxy of 4 times the global number density, eq. (5) also indicates that for these high redshift galaxies, the bias factor is quite large, $\gtrsim 3$.

**(2) Tidal Force.** If as some models claimed, that Ly$\alpha$

Table 3. Lyα line density in the neighborhood of metal systems in spectra of three close quasars.

| QSO | Lyα line density | Metal system | nearby Lyα line density in | | |
|---|---|---|---|---|---|
| | | | 1628+2651A | 1628+2653 | 1628+2651B |
| 1628+2651A | 30/(495Å) | z=2.275 | – | 0 | 0 |
| 1628+2653 | 38/(597Å) | z=2.053 | 0 | – | 0 |
| | | z=2.401 | 1/(6.8Å) | – | – |
| 1628+2651B | 18/(339Å) | z=1.987 | – | 1/(5.1Å) | – |
| | | z=2.095 | 2/(5.4Å) | 0 | – |
| | | z=2.241 | 2/(5.9Å) | 2/(5.9Å) | – |

clouds are confined by dark matter mini-halos, the clouds can be tidally disrupted in a strong gravitational field. Even if gravity is only part of the total confining force, we can express the confining force by gravity multiplied by a factor, $1/\epsilon$. For a Lyα cloud at a physical distance of $1h^{-1}$ Mpc from a group of galaxies, the cloud will be tidally disrupted if

$$\frac{GM}{R^2}\frac{r}{R} > \frac{Gm}{\epsilon r^2} \approx \frac{4\pi}{3}\frac{G\rho_{\rm DM} r}{\epsilon} \quad (6)$$

up to a numerical factor of order 1. $M$ is the mass within a radius of $R$. $m$ is the mass of the cloud, and is roughly $4\pi\rho_{\rm DM} r^3/3$ where $\rho_{\rm DM}$ is the average dark matter density within the cloud. $R = 1h^{-1}$ Mpc in our consideration. $r$ is the size of the cloud, typically in the range from $60h^{-1}$ kpc to $500h^{-1}$ kpc (Fang et al. 1995). Thus to stablize a Lyα cloud against tidal force requires

$$M \lesssim \frac{4\pi}{3}\rho_{\rm DM} R^3/\epsilon \quad (7)$$

Since Lyα clouds with $10^{13}\rm cm^{-2} < N_{\rm HI} < 10^{14}\rm cm^{-2}$ are structures not yet completely collapsed, their density fluctuation $\delta\rho/\rho \sim 1$, eq. (7) is simply

$$(\delta M/M)_{1h^{-1}\rm Mpc} \lesssim 1/\epsilon. \quad (8)$$

The equation is therefore a relaxed version of eq. (5).

(3) **Photoionization.** It has been shown that the ionization photon flux from galaxies can be comparable to that from quasars (Madau & Shull 1995, and references therein). The flux experienced by a Lyα cloud near a group of galaxies clustering around a detected galaxy, therefore, can be significantly higher than the background, since the local group of galaxies has higher galaxy number density than the average. Calculating the flux, however, may be difficult because it depends on the positions of the Lyα cloud and the galaxies and is quite likely dominated by its immediate neighbors. Given that the Lyman continuum photon luminosity from a typical disk galaxy is $\sim 10^{42}$ erg·s$^{-1}$, a very naive estimate shows that 15 such galaxies $1h^{-1}$ Mpc away yield a ionizing photon flux of $\sim 5\times 10^{-22} f$ erg·cm$^{-1}$·s$^{-1}$·Hz$^{-1}$ at 1 Ryd, where $f$ is the fraction of Lyman continuum radiation that escapes the galaxy. If $f > 0.1$, the flux contributed by the local group of galaxies can be a non-negligible contribution to background ionizing photon flux of $\sim 10^{-21}$ erg·cm$^{-1}$·s$^{-1}$·Hz$^{-1}$ (Bajtlik et al. 1988; Lu et al. 1991).

The column density of a Lyα cloud scales inversely with the local ionization photon flux. Given a local ionization photon flux that is $1 + \omega$ times the background flux, the ratio of the local Lyα line density at $1h^{-1}$ Mpc away from a detected galaxy to the average Lyα line density, assuming $dN/dN_{\rm HI} \propto N_{\rm HI}^{1.7}$ (with $N$ being the Lyα line density) (Bajtlik 1993), is

$$\frac{\int_a^b (dN/dN_{\rm HI})_{1h^{-1}\rm Mpc} dN_{\rm HI}}{\int_a^b \langle dN/dN_{\rm HI}\rangle dN_{\rm HI}} = \frac{\int_{a/(1+\omega)}^{b/(1+\omega)} N_{\rm HI}^{1.7} dN_{\rm HI}}{\int_a^b N_{\rm HI}^\beta dN_{\rm HI}} = \left(\frac{1}{1+\omega}\right)^{0.7}. \quad(9)$$

Data in section 2 preclude a strong proximity effect at $1h^{-1}$ Mpc distance, thus $\omega$ can't be significantly larger than 1. This is not surprising based on our naive calculation of the local ionization photon flux.

(4) **Pressure confinement.** If Lyα clouds in our $N_{\rm HI}$ range are mainly confined by pressure, either from IGM or shock waves, the clouds will not remain stable if the gravity of neighboring galaxies overcome the pressure gradient $dP/dR$ inside the cloud, i.e.,

$$\frac{GM}{R^2} > \frac{2kT}{\mu_{\rm H}}\frac{d\ln P}{dR} \quad (10)$$

where $M$ is the mass of the group of galaxies within $1h^{-1}$ Mpc of the detected galaxy, $T$ is the temperature of the cloud, and $\mu_{\rm H}$ the atomic weight of hydrogen. An equation of state for ideal (ionized) gas is assumed. Given $T \sim 3\times 10^4$ K, stablization of these Lyα clouds requires

$$M \lesssim \frac{2kTR^2}{G\mu_{\rm H}}\frac{d\ln P}{dR} \sim 10^{12} M_\odot \left(\frac{120 h^{-1}\rm kpc}{h_{\rm p}}\right) \quad (11)$$

where $h_{\rm p}$ is the pressure scale height $d\ln P/dR$. On average, $h_{\rm p}$ is expected to be roughly about or larger than the line of sight dimension of the clouds. Since we know that there are $1.5\times 10^{13} M_\odot$ (with $\sim 15$ galaxies each weighing $\sim 10^{12} M_\odot$) to $10^{14} M_\odot$ ($\delta M/M \lesssim 1$) excessive mass within $1h^{-1}$ Mpc radius, the line of sight dimension has to be $\lesssim 1$–$8h^{-1}$ kpc, more than an order of magnitude smaller than their transverse size. In other words, clouds confined by pressure have to be filaments or sheets.

Clearly an investigation of the proximity effect of galaxies down to $1h^{-1}$ Mpc is not enough. Much more information can be obtained if this effect is reliably determined at a closer distance, such as $\lesssim 0.5h^{-1}$ Mpc. Then the implication on gravitational potential may be directly compared with predictions of cosmological models. If there exists clear deficiency in number of Lyα clouds (without contamination from galaxies) at $\lesssim 0.5h^{-1}$ Mpc, one can also explore how ionization flux from galaxies, tidal force, mergers may play in destroying them. Current data show that the number density of Ly$\alpha$ clouds with $13.3 \leq \log N_{\rm HI} < 13.8$ within $\sim 0.6 h^{-1}$ Mpc of a galaxy is below the average Ly$\alpha$ line density, but its error is about $\sim 40\%$. Thus, to narrow the error bar to $\sim 20\%$ requires high resolution data from at least a dozen more high redshift quasars. It will also be very useful to probe the proximity effect on Ly$\alpha$ clouds with different $N_{\rm HI}$, which is not fully explored in our limited sample.

Numerical simulations of the proximity effect of galaxies can also provide very valuable information. Current simulation of Ly$\alpha$ clouds may not have sufficient dynamic range to simulate both galaxy formation (and perhaps galaxy cluster formation) and Ly$\alpha$ cloud formation simultaneously. Future simulations of all of these objects and their interplay will certainly be very interesting.

In summary, the proximate effect of metal line absorption systems in quasar spectra is explored. We find that for Ly$\alpha$ absorption lines with $13.3 \leq \log N_{\rm HI} < 13.8$, there is no strong proximity effect down to $\sim 500$ km/sec away from metal line absorption systems at $z \sim 3$. At velocity splits below 500 km/sec, Ly$\alpha$ absorption lines with $13.3 \leq \log N_{\rm HI} < 13.8$ may suggest a deficit in number of lines. Ly$\alpha$ absorption lines with higher HI column densities show a possible clustering around metal line absorption systems because of the inclusion of Ly$\alpha$ absorption arising from these metal line absorption systems. A Ly$\alpha$ absorption sample with $N_{\rm HI} \gtrsim 10^{13.3}$ cm$^{-2}$ in spectra of three quasar pairs (whose lines of sight are separated by $\sim 0.5 h^{-1}$ Mpc) is also investigated for the proximity effect but its statistics is too low to draw a reliable conclusion. Possible implications of our finding is discussed. More information on early structure formation can be extracted if the proximity effect is reliably determined below $\sim 500$ km/sec. Further observations that can provide better statistics and remove the contamination due to metal line absorption systems in Ly$\alpha$ absorption samples are suggested.

## 4 ACKNOWLEDGEMENT

The author thanks Man-hoi Lee, Avery Meiksin and Larry Widrow for helpful discussions. This work is supported by NSERC grant at Queen's University, and by CITA National Fellowship at Canadian Center for Theoretical Astrophysics.